\documentclass[aip,rsi,amsmath,amssymb,reprint,]{revtex4-1}

\usepackage{graphicx}
\usepackage{dcolumn}
\usepackage{bm}
\usepackage{mathptmx, textcomp}
\usepackage{amsmath}
\usepackage{amssymb}

\begin{document}

\title{An apparatus for immersing trapped ions into an ultracold gas of neutral atoms}

\author{Stefan Schmid}
  \affiliation{Institut f\"ur Quantenmaterie und Center for Integrated Quantum Science and Technology \textit{IQST}, Universit\"at Ulm, 89069 Ulm, Germany}
 \affiliation{Institut f\"ur Experimentalphysik und Zentrum f\"ur Quantenphysik, Universit\"at Innsbruck,
 6020 Innsbruck, Austria}

\author{Arne H\"arter}
  \affiliation{Institut f\"ur Quantenmaterie und Center for Integrated Quantum Science and Technology \textit{IQST}, Universit\"at Ulm, 89069 Ulm, Germany}
 \affiliation{Institut f\"ur Experimentalphysik und Zentrum f\"ur Quantenphysik, Universit\"at Innsbruck,
 6020 Innsbruck, Austria}

\author{Albert Frisch}
 \affiliation{Institut f\"ur Experimentalphysik und Zentrum f\"ur Quantenphysik, Universit\"at Innsbruck,
 6020 Innsbruck, Austria}

\author{Sascha Hoinka}
 \altaffiliation[Present address:]{ARC Centre of Excellence for Quantum-Atom Optics, Centre for Atom Optics and
 Ultrafast Spectroscopy, Swinburne University of Technology, Melbourne 3122, Australia}
 \affiliation{Institut f\"ur Experimentalphysik und Zentrum f\"ur Quantenphysik, Universit\"at Innsbruck,
 6020 Innsbruck, Austria}

\author{Johannes Hecker Denschlag}
 \affiliation{Institut f\"ur Quantenmaterie und Center for Integrated Quantum Science and Technology \textit{IQST}, Universit\"at Ulm, 89069 Ulm, Germany}
 \affiliation{Institut f\"ur Experimentalphysik und Zentrum f\"ur Quantenphysik, Universit\"at Innsbruck,
 6020 Innsbruck, Austria}
 \email{johannes.denschlag@uni-ulm.de}

\begin{abstract}
We describe a hybrid vacuum system in which a single ion or a well
defined small number of trapped ions (in our case Ba$^+$ or
Rb$^+$) can be immersed into a cloud of ultracold neutral atoms
(in our case Rb). This apparatus allows for the study of
collisions and interactions between atoms and ions in the
ultracold regime. Our setup is a combination of a Bose-Einstein
condensation (BEC) apparatus and a linear Paul trap. The main
design feature of the apparatus is to first separate the
production locations for the ion and the ultracold atoms and then
to bring the two species together. This scheme has advantages in
terms of stability and available access to the region where the
atom-ion collision experiments are carried out. The ion and the
atoms are brought together using a moving 1-dimensional optical
lattice transport which vertically lifts the atomic sample over a
distance of 30$\,$cm from its production chamber into the center
of the Paul trap in another chamber. We present techniques to
detect and control the relative position between the ion and the
atom cloud.
\end{abstract}


\maketitle

\section{Introduction}
In recent years, both the fields of cold trapped ions and of
neutral,  ultracold atomic gases have experienced an astonishing
development. Full control has been gained over the respective
systems down to the quantum level. Single ions can be selectively
addressed and their quantum states can be coherently manipulated
and read out\cite{Lei2003}. The collective behavior of neutral
atomic quantum gases can be mastered by controlling the
particle-particle interactions, temperature, and physical
environment\cite{BEC2002}.

Over the last two decades increasing  efforts have been made to
study cold collisions between ions and neutral particles. One
approach is to study collisions in a cold He buffer gas (see for
example.\cite{Haw1991,Ger2002,Ger2005,Ott08}) Another approach for
collisions in the regime of a few K uses neutral,
velocity-selected particles from a beam of molecules to collide
with trapped ions.\cite{Wil2008} In recent years collisions
between atoms in a magneto-optical trap (MOT) and trapped ions
trap have been
observed.\cite{Cet2007,Gri2009,Hal11,Rel11,Rav11,WeidWest} In
2010, in parallel to the group of M. K\"ohl
\cite{Zip2010,Zip2010a,Zip2011}, our group has finally
demonstrated immersing trapped ions in a BEC of Rb atoms at nK
temperatures.\cite{Sch2010a}

\begin{figure*}
\includegraphics[width=15cm]{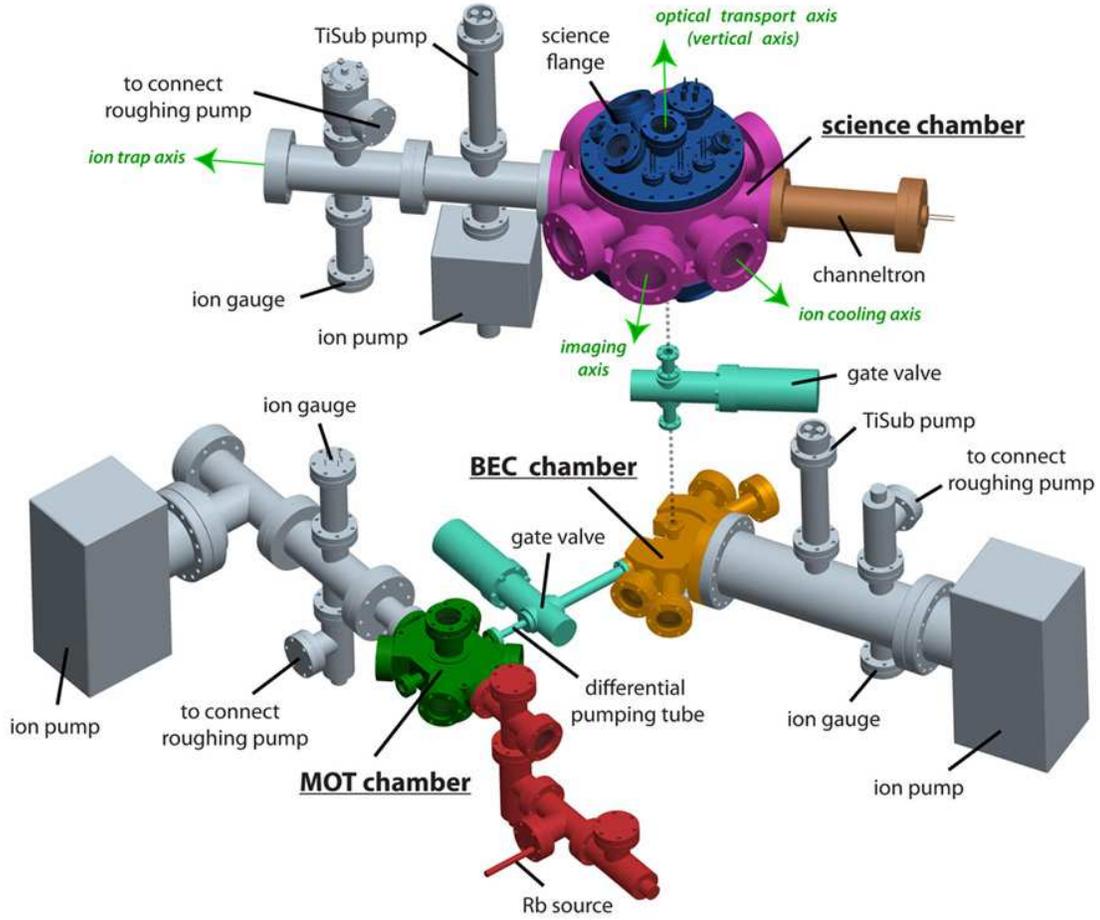}
\caption{Layout of the vacuum apparatus in a partially exploded
view: The science chamber (upper section) is connected to the BEC
chamber (in the lower section) via a differential pumping stage (turquoise)
along the dashed vertical axis. The MOT chamber (green) and the BEC chamber (red) which form the lower section are also connected via a differential pumping stage (turquoise). The science
chamber exhibits a large DN200CF flange (blue) on top, the
``science flange", onto which the ion trap (not shown here) is
mounted. A channeltron detector
(brown) is connected along the axis of the linear ion trap. All three chambers are evacuated by their own pumping sections (grey).
(Note: For better visibility the upper section is rotated
clockwise by 90$^\circ$ around the dashed vertical axis.)}
 \label{vacuumsysoverview}
\end{figure*}

Here we describe the hybrid apparatus used for our atom-ion collision experiments in detail. A central design
concept of our setup is the spatial separation of the BEC
apparatus - where the ultracold atoms (or a BEC) are produced - from
the ion-trapping region, where the atom-ion collision experiments
are performed. This way we gain valuable optical access to the
atom-ion interaction region, that is necessary for trapping,
manipulating, and detecting the atoms and ions. Furthermore, the
separation and isolation of the production sites ensures that any
mutual disturbance between the radiofrequency (rf) Paul trap and
the rf used for forced evaporative cooling of the atomic sample is minimized. To transport the atoms over 30$\,$cm from their
place of production to the trapped ions, we employ a
moving 1-dimensional (1-d) optical lattice.

As demonstrated in a first set of experiments \cite{Sch2010a} the
apparatus enables us to study elastic and inelastic atom-ion
collisions in the ultracold regime.\cite{Cot2000b, Idz2009} We
plan to investigate cold chemical reactions and the controlled
formation of cold molecular ions, topics which recently gained
considerable interest (see for example \cite{Sta2010,Sch2010b}).
Furthermore, the apparatus allows for carrying out other
interesting lines of research. There are proposals to study the
dynamics of charged impurities in a quantum degenerate
gas,\cite{Mas2005,Kal2006,Cuc2006} charge transport in a gas in
the ultracold domain,\cite{Cot2000a} or the formation of a
mesoscopic molecular ion.\cite{Cot2002}

The article is organized as follows:  Section~II describes the
layout of the multi-chamber vacuum apparatus. In Sec.~III, the
design and the operation of the ion trap are discussed. Section~IV
addresses the preparation of the ultracold atom cloud in the BEC
apparatus and its optical transport into the science chamber, in which the ion trap is located. In
particular, we describe an experimental procedure based on atom-ion
collisions used to precisely position the atom cloud with respect
to the ion.

\section{The vacuum apparatus}

\begin{figure}
\includegraphics[width=8.5cm]{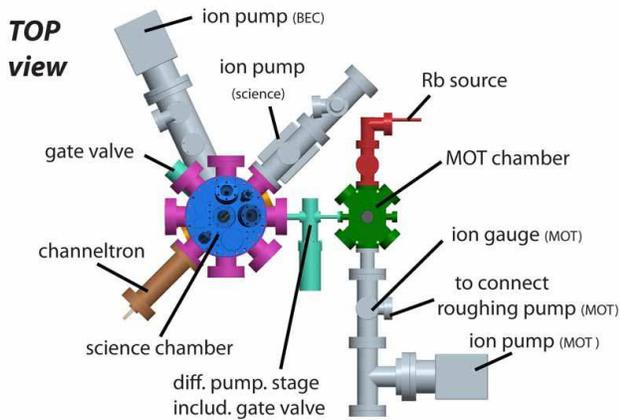}
\caption{Top view of the vacuum system: The science chamber (red),
covers up the BEC chamber (not visible).}
\label{vacuumsysoverview2}
\end{figure}

The vacuum apparatus consists of three main building blocks.  A
MOT chamber for trapping and laser cooling Rb atoms. A BEC chamber
for evaporative cooling of the Rb atoms. A ``science chamber" that
houses the ion trap and where the ion-atom collision experiments
take place. The apparatus has two floors: MOT chamber and BEC
chamber form the lower section. The science chamber, forming the
upper section,  is located 30$\,$cm above the BEC chamber (see
Figure\,\ref{vacuumsysoverview} and \ref{vacuumsysoverview2}).
 The three vacuum chambers are connected via two differential pumping
stages, each consisting of a differential pumping tube as well as
a gate valve. If necessary, the vacuum chambers can be separated
from each other by closing the gate valves. A series of vacuum
gauges, pumps, and valves is used to evacuate the system and to
determine the pressure. By baking out the setup at temperatures
between 180$^{\circ}$C and 250$^{\circ}$C, ultrahigh vacuum (UHV)
conditions are established in all three chambers. When in
operation, the pressures are approximately 10$^{-9}\,$mbar in
the MOT chamber and 10$^{-11}\,$mbar in the BEC chamber and the science chamber.

\subsection{Lower section: BEC apparatus}

The stainless steel (AISI 316L) MOT chamber features ten optical
viewports, which are required to implement the MOT laser beams, to
connect to the Rb vapor source, to pump the chamber and to move
the atoms out of the MOT chamber towards the BEC chamber. An ion
getter pump\cite{igp} is used to keep the MOT chamber at UHV
conditions, as measured by an UHV pressure gauge (Bayard Alpert
type).\cite{gauge}

The Rb vapor source is an ampule filled with bulk Rb and He as an
inert gas \cite{Rb}. Since Rb is very reactive when exposed to
air, the ampule is not cracked until the surrounding ``oven
section" has been evacuated. Once Rb has been released from the
ampule, the pressure in the oven section is determined by the Rb
vapor pressure, which is 4$\times$10$^{-7}\,$mbar at room
temperature. As a consequence, the pressure in the center of the
MOT chamber increases from its original value of 10$^{-11}\,$mbar
to a few times 10$^{-9}\,$mbar and is then completely dominated by
the Rb vapor. If necessary, the vapor pressure can be adjusted by
heating the Rb source or by changing the setting of the valve
which separates the oven section from the MOT chamber.

The pressure in the BEC chamber is below 10$^{-11}\,$mbar using a
combination of a titanium sublimation (TiSub) pump \cite{TiSub}
and an ion getter pump \cite{igp}. At this pressure we achieve
lifetimes of the atom cloud of more than 2$\,$min which is
sufficient to carry out rf evaporative cooling.

In order to maintain a pressure gradient of
$p_\textsc{mot}/p_\textsc{bec} \approx 10^2$, a differential
pumping tube is used to separate the MOT and the BEC chamber.
Since the size of the atom cloud, which has to be transported
through the tube, amounts to a few millimeters, we chose the tube
diameter to be 8$\,$mm. Molecular flow calculations then determine
the tube length to be 115$\,$mm.

\begin{figure}
\includegraphics[width=8.0cm]{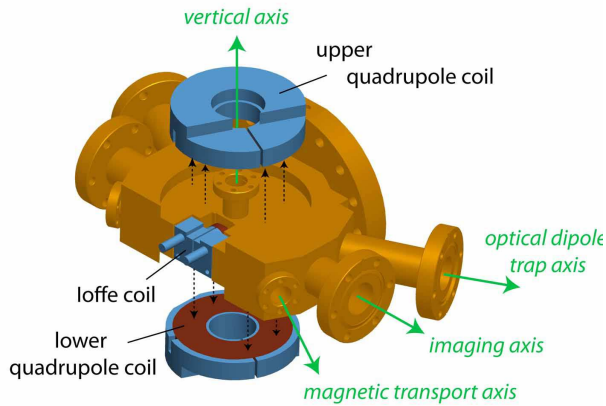}
\caption{Explosion view of the BEC chamber and the magnetic QUIC
trap coils. The QUIC trap is generated by the two quadrupole coils and the Ioffe coil (blue). The coils are mounted outside the vacuum
to the walls of the BEC chamber. The atoms enter the chamber along
the magnetic transport axis and leave it along the vertical axis.}
\label{becchamber}
\end{figure}

\begin{figure}
\includegraphics[width=7cm]{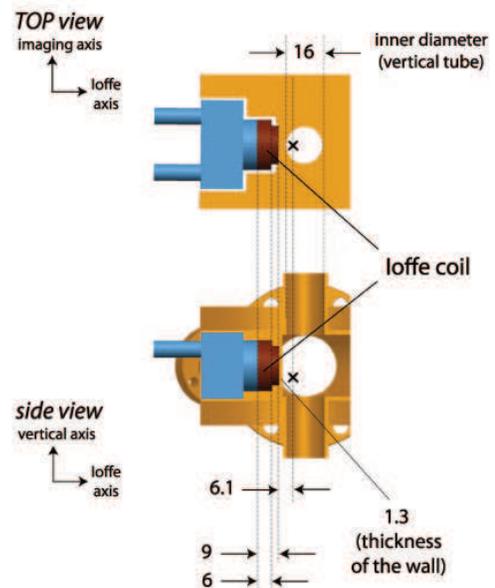}
\caption{BEC chamber and Ioffe coil: The BEC chamber features a small insertion slot with an end wall thickness of only 1.3$\,$mm, so that the Ioffe coil (brown) and its holder (blue) can be mounted at a minimum distance of only 11.1$\,$mm from the center of the chamber. The position of the atom cloud, when it is stored in the QUIC trap, is denoted by the black cross. The dimensions are given in mm.}
\label{becchamberdetail}
\end{figure}

The design of the BEC chamber is shown in
Fig.$\,$\ref{becchamber} and \ref{becchamberdetail}. After laser cooled atoms from the MOT
are magnetically transported into the BEC chamber, evaporation in
a Quadrupole-Ioffe configuration (QUIC) trap \cite{Ess1998} brings
the atoms to BEC or close to BEC. Afterwards they are vertically
transferred to the ion trap in the science chamber. QUIC traps are typically used in combination with a glass cell, as the Ioffe coils need to be placed quite close to the atoms. In our setup where the BEC chamber is physically connected along the horizontal (to the MOT chamber) and the vertical (to the science chamber) direction, strong shear forces are acting on the chamber walls. For this reason we
chose stainless steel instead of glass for the construction of the BEC chamber. To minimize the distance between the atoms and the Ioffe coil, the BEC chamber exhibits a special insertion slot with a thin end
wall into which the Ioffe coil can be placed (see Fig.$\,$\ref{becchamberdetail}). The BEC chamber features four optical axes. The
first axis points along the vertical direction and is needed for
the optical transport of atoms from the BEC chamber into the
science chamber (transport axis). On both ends small DN16CF
flanges are used, so that the QUIC quadrupole coils can be easily
mounted. The second axis (imaging axis) is used for imaging the
ultracold atoms. DN40CF viewports are used on both ends, in order
to allow for a good imaging quality. In addition, optical access
along the magnetic transport axis is desirable, in order to be
able to image the atom cloud at any intermediate position of the
transport. Finally, the BEC chamber features a fourth optical axis
(optical dipole trap axis), which is
currently not used in our experiments, but could for example be
employed for the addition of a dipole trapping beam. In order to utilize the
full pumping speed of our ion pump, the pumping section is
connected to the BEC chamber via a DN100CF flange.

\subsection{Upper section: Science chamber}

\begin{figure}
\includegraphics[width=8.5cm]{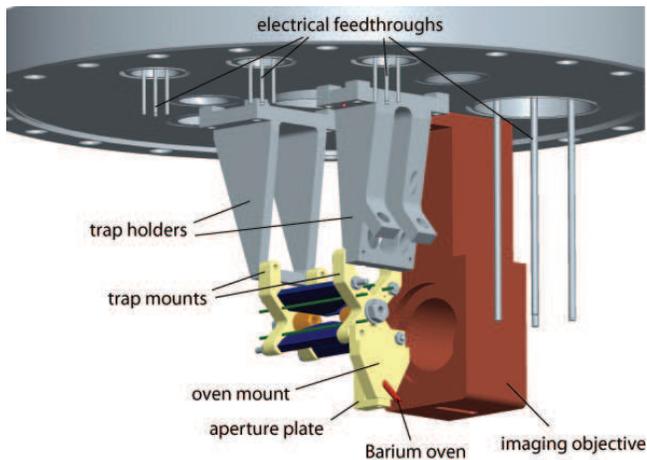}
\caption{Science flange (DN200CF): The Paul trap (blue and golden) as well as the Ba oven (red) are mounted on MACOR ceramic parts (pale
yellow). The imaging objective consists of four lenses, all of them held in place by a massive aluminum mount (brown) (see also section
\ref{imaging}).}
\label{scienceflange}
\end{figure}

The science chamber (Fig.$\,$\ref{vacuumsysoverview} and \ref{vacuumsysoverview2}),
represents the heart of our vacuum apparatus since this is where
the experiments take place.  It is designed for maximum optical
access with eight optical axes. The optical access is needed for
cooling and imaging of the ions as well as for trapping,
manipulating, and imaging of the atoms. All parts of the ion trap
as well as an object lens to collect the ion fluorescence are
mounted within the science chamber onto the ``science flange" (Fig.$\,$\ref{scienceflange}).
The flange features various electrical feedthroughs, which are needed
to apply the required high voltages to the Paul-trap electrodes
and to run currents of up to 12$\,$A through the Ba oven. The
DN200CF science flange is mounted on top of the science chamber.

The stainless steel, octagon-shaped science chamber
(Fig.$\,$\ref{vacuumsysoverview})  is evacuated by a combination
of an ion getter pump and a TiSub pump. As an optional addition of
our setup, we have connected a channeltron ion detector
\cite{channel} along the axis of the linear ion trap (see also
Fig.$\,$\ref{vacuumsysoverview} and \ref{vacuumsysoverview2}). One
possible application of the channeltron is the identification of
ions via time-of-flight mass spectrometry as has been demonstrated
in other experiments (see for example \cite{Rav2011}).

\section{Ion trapping}

\subsection{Linear Paul trap}

\begin{figure}
\includegraphics[width=7cm]{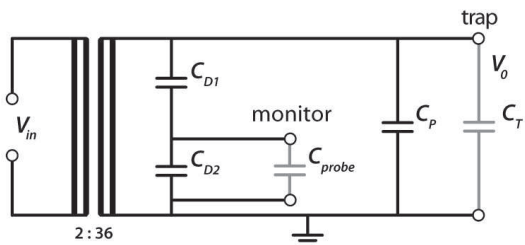}
\caption{Ferrite-toroid transformer with a turns ratio of 2:36. By adjusting the capacity $C_\textrm{P}$  the impedance of the trap (corresponding to $C_\textrm{T}$) is matched and the ratio between the output and the input voltage ($V_\textrm{O} / V_\textrm{in}$) is maximized. In order to be able to monitor the output voltage, a capacitive voltage divider ($C_\textrm{D1}$ and $C_\textrm{D2}$) is used.}
\label{Trafo}
\end{figure}

We employ a linear radiofrequency (rf) Paul trap\cite{Rai1992}
(see Fig.$\,$\ref{iontrap}) to store Ba$^+$ or Rb$^+$ ions. For
the trapping of the ions along the two radial directions we use
four blade electrodes, which are mounted symmetrically at a
distance of $r_0 = 2.1\,$mm from the trap axis. A rf of $\Omega =
2\pi\times$5.24$\,$MHz with an amplitude of 2$\times
U_\textrm{rf}$ = 1400$\,$V$_\textrm{pp}$ is applied to two of the
four blades, whereas the other two are grounded. The rf is
generated by a commercially available function generator
\cite{Agilent} and subsequently amplified by a 5$\,$W rf amplifier
from Minicircuits\cite{Minicircuits5W}. The power at the output of
the amplifier is inductively coupled to the trap electrodes via a
ferrite-toroid transformer (see Fig.$\,$\ref{Trafo}). The
impedance of the trap is matched, so that the supply voltage is
resonantly enhanced by a factor of 30. Trapping along the axial
direction is achieved by applying dc voltages on the order of
100$\,$V to the two endcap electrodes which are located on the
trap axis at a distance of 7$\,$mm from the trap center. Typical
ion trapping frequencies for the parameters given above are
$\omega_{\textrm{\,rad}} \approx 2\pi\times$250$\,$kHz  and
$\omega_{\textrm{\,ax}} \approx 2\pi\times$80$\,$kHz for Ba$^+$
and $\omega_{\textrm{\,rad}} \approx 2\pi\times$390$\,$kHz and
$\omega_{\textrm{\,ax}} \approx 2\pi\times$100$\,$kHz for Rb$^+$.
As expected, the trap frequencies of our ions in the linear Paul
trap scale in first order as $\omega_{\textrm{\,rad}} \propto 1/m$
and $\omega_{\textrm{\,ax}} \propto \sqrt{1/m}$. The stability
factor $q$ is generally given by $q = 2eU_\textrm{rf}/(m \gamma
r_0^2 \Omega^2)$\cite{Ber1998}, where $\gamma = 1.53$ is a
numerical factor that depends on the exact geometry of the rf
electrodes. Our Paul trap allows stable trapping of both, Ba$^+$
and Rb$^+$, since $q\ll 1$ for both species ($q=0.13$ for Ba$^+$
and $q=0.21$ for Rb$^+$).

All electrodes are made of non-magnetic, high-grade, stainless steel
of type AISI 316L. This material is specified to have a magnetic
permeability of $\mu < 1.005$. The blades are produced by
electrical discharge machining, which allows for a higher
precision and a smaller surface roughness as compared to milling.
The electrodes are mounted onto two insulating parts, which are
made out of a machinable glass-ceramic (MACOR). This material has
a very low outgassing rate and is thus well suited for UHV
applications.

\subsection{Loading and laser cooling of ions}

\begin{figure}
\includegraphics[width=8cm]{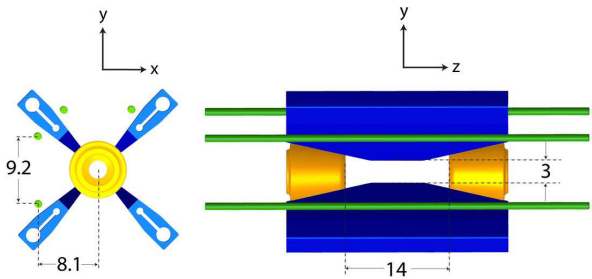}
\caption{Linear ion trap (Paul trap). The
trap consists of four rf electrodes (blue) for confinement in the
x-y plane, two endcap electrodes (golden) for confinement along
the z axis, and four compensation electrodes (green) for the
generation of dc electrical fields in the x-y plane. All
electrodes are made of stainless steel AISI 316L. The dimensions are given in mm.} \label{iontrap}
\end{figure}

We load the Paul trap (Fig. \ref{iontrap}) with either
$^{138}$Ba$^{+}$ or $^{87}$Rb$^{+}$ ions. To work with Ba$^{+}$,
we run a current of about 8$\,$A (corresponding to 6$\,$W) through
the commercially available Ba source\cite{Baoven} (see Fig.$\,$\ref{scienceflange}). It is a
stainless steel tube with a diameter of 2$\,$mm, which is filled
with metal alloy containing Ba. Ba vapor is created through
sublimation out of the alloy.  In the center of the trap, the
neutral Ba atoms emitted by the oven are photoionized using a
diode laser operating at 413$\,$nm.\cite{413er} A few mW of laser
power are used to drive the resonant two-photon transition from
the ground state to the continuum via the $^3$D$_1$ state. With
this procedure we are able to load single ions into our trap within a few minutes.

We perform Doppler cooling of the $^{138}$Ba$^{+}$ ions  on the
6S$_{1/2} \rightarrow$6P$_{1/2}$ cycling transition, which has a
transition wavelength of 493$\,$nm and a linewidth of 15.1$\,$MHz.
The corresponding Doppler temperature is 360$\,$\textmu K. The
493$\,$nm cooling light is generated via frequency doubling of a
986$\,$nm diode laser. Both the diode laser and the
frequency-doubling stage are part of the commercially available
system ``DL SHG" from Toptica. To stabilize the frequency of the
493$\,$nm light, the 986$\,$nm laser is locked to a temperature-stabilized optical
cavity using the well-establishd Pound-Drever-Hall method\cite{Dre1983,Bla2001}.
An additional ``repumper" laser at a wavelength of 650$\,$nm is
needed, in order to bring the atoms from the metastable 5D$_{3/2}$
state back into the cooling cycle. The repumper is a home-made
external-cavity diode laser, which features an
anti-reflection-coated laser diode to guarantee stable lasing at
the desired wavelength. The frequency of the repumper is
stabilized using the same locking scheme as for the 986$\,$nm
laser.

To work with clouds of Rb$^{+}$ ions, we transport ultracold Rb
atoms  into the center of the Paul trap and ionize them using the
imaging laser at 780$\,$nm together with the ionization laser at
413$\,$nm. The resonant imaging laser brings the Rb atom into the
5P$_{3/2}$ state, so that the 413$\,$nm laser can then drive the
transition into the continuum. Since we start with an ultracold
trapped atom source, this ionization procedure is very fast and
efficient. It allows for loading clouds of Rb$^{+}$ ions into the
Paul trap within a few milliseconds.

We employ a different scheme when performing experiments  with
single Rb$^{+}$ ions. In this case, we first load a single
Ba$^{+}$ ion into the Paul trap following the procedure described
above. Then we let the Ba$^{+}$ ion interact with ultracold Rb
atoms until the charge transfer process
Rb$\,$+$\,$Ba$^+\rightarrow\,$Rb$^+$+Ba$\,$ has taken place. This
takes typically a few seconds. The newly formed Rb$^+$ ion is
``dark", as it cannot be detected via standard fluorescence
imaging due to the lack of an accessible cycling transition.
Therefore Rb$^+$ has to be detected via its elastic collisions
with the neutral atoms and the corresponding atom losses. The
newly formed Rb$^+$ ion is available for thousands of experimental
cycles, since its lifetime in the Paul trap is typically on the
order of days.

\subsection{Micromotion}

In addition to the pseudopotential generated by the rf drive,
dc stray electric fields are also present. Possible sources
of these fields are imperfections in the fabrication of the trap
or patch charges on the ceramic parts. Such surface charges could
be generated by our blue Ba$^+$ lasers (413$\,$nm and 493$\,$nm)
via the photoelectric effect. In any case, the dc fields push the
ion out of its ideal trap location, the rf node,  into a region of
non-vanishing rf, leading to the so-called ``excess micromotion"
of the ion.\cite{Ber1998} In
order to minimize this enhanced micromotion, we have to compensate
the dc electric offset fields at the position of the ion. To
compensate electrical offset fields in the axial direction we can
apply corresponding voltages to the endcaps of the Paul trap. To
generate compensation fields along the radial direction, four
``compensation electrodes" (two for each direction) are added to
the design of the linear Paul trap. By placing the electrodes at a
distance of 9.35$\,$mm from the trap axis and applying a voltage
$U_\textrm{comp}$, we are able to generate dc compensation fields
of $E_\textrm{comp}=\beta\,U_\textrm{comp}$, where
$\beta=3.1\,$m$^{-1}$.

A simple and in our system very accurate method to detect excess micromotion is to
determine the position shift of the ion when the rf amplitude is
changed. By adjusting the compensation fields such that the
position shift is minimized, we can assure that the potential
minimum nearly coincides with the rf node. This method works
nicely for electric fields along the vertical direction. However,
for the compensation of fields along the $x$-axis a different
method has to used, since we are not able to measure the $x$
position of the ion with high accuracy. One possibility, realized
in our setup, is to modulate the amplitude of the rf drive. Setting the modulation frequency equal to the trap frequency
leads to resonant heating of the ion, which can be detected via a
smearing of the ion fluorescence. The heating is particularly
strong when the ion is not in the rf node. Hence, we can adjust
the compensation fields by minimizing the heating. In early
experiments, using the methods described here, we were able  to
reduce the dc electric fields at the position of the ion to about
1$\,$V/m.\cite{Sch2010a}

\subsection{Imaging methods}
\label{imaging} We detect the ion by collecting its fluorescence
using a high-aperture laser objective (HALO) from Linos (see Fig.
\ref{iontrap}). The HALO has a numerical aperture of
NA$\,=\,$0.2 and a focal length of $f=60\,$mm, which enables us to
detect about NA$^2/4 \approx 1\%$ of the spontaneously emitted
photons. It is placed inside the vacuum chamber at a distance of
$f = 60\,$mm from the trap center. Since the original mount of the
HALO is anodized and generally not designed to be put into a UHV
environment, it was exchanged by a UHV-capable aluminum mount.
This new mount features an air vent in order to avoid slow
outgassing of air that is enclosed in between the different lenses
of the objective.

The collimated fluorescence light exits the chamber through  a
DN63CF AR-coated viewport. An $f=300\,$mm achromat is then used to
focus the light onto the EM-CCD chip of an Andor Luca(S) camera.
Diffraction at the aperture of the HALO ultimately limits the
resolution of the imaging system to about 1.5$\,$\textmu m, which
is an order of magnitude smaller than the typical distance between
two neighboring ions of an ion string.

Absorption imaging of the cloud of neutral atoms is also done with
the HALO. To separate the Rb imaging beam (780$\,$nm) from the
Ba$^+$ fluorescence light (493$\,$nm), we use a dichroic mirror.
Together with the HALO, an $f=200\,$mm achromat forms the
objective for the neutral atom detection. The large spacing
between the ion-trap electrodes allows for taking the images not only in-situ
but also after a free expansion of the atom cloud of up to 15$\,$ms.

\section{Preparation and delivery of ultracold atomic samples}

After the ion has been trapped, we start the production of an
ultracold atom cloud in the lower section of the vacuum apparatus.
The time needed to create a Rb cloud with 2$\times$10$^6$ atoms at
a temperature of about 1$\,$\textmu K is approximately 35$\,$s.
Another 5$\,$s are required to transport the atom cloud into the
science chamber and to perform further forced evaporative cooling
in an optical dipole trap down to BEC or to cold thermal
ensembles with typical temperatures of 100$\,$nK.

\subsection{MOT loading and magnetic trap}

To operate the MOT, we have set up two external-cavity diode
lasers tuned to the 5$^2$S$_{1/2}\rightarrow$5$^2$P$_{3/2}$
transition in $^{87}$Rb. One of the diode lasers is locked to the
$|\,F=2\rangle \rightarrow |\,F'=3\rangle$ cycling transition
using the modulation transfer spectroscopy
technique.\cite{McC2008} The light from this laser is amplified
with a tapered amplifier \cite{BoosTA} and sent through a polarization-maintaining (PM) optical fiber. After the fiber the total power of
250$\,$mW is split up into six beams, all of them having a diameter of 30$\,$mm.
Using acousto-optical modulators the cooling light is detuned to
about -3.5$\Gamma$ relative to the cycling transition, where
$\Gamma=6\,$MHz is the transition linewidth. Our second Rb laser
is locked to the $|\,F=1\rangle \rightarrow |\,F'=1\rangle /
|\,F'=2\rangle$ cross over line using the frequency modulation
(FM) technique.\cite{Bjo1983} This repumper laser is used to pump
the atoms from the $|\,F=1\rangle$ groundstate back into the
cycling transition. To operate the MOT a total repump power of
6$\times$1.5$\,$mW is employed. The required magnetic field
gradient of $B_z' = 8\,$G/cm is generated by running a current of
5$\,$A through a pair of anti-Helmholtz coils. This setup enables
us to load about 3$\times$10$^9$ $^{87}$Rb atoms from the
background vapor into the MOT.

By turning off the magnetic field and detuning the  MOT cooling
beams to about -8.5$\,\Gamma$, we perform polarization-gradient
cooling for a duration of 10$\,$ms. In this way, the temperature
of the atoms is reduced to about 40$\,$\textmu K. In a next step
the atoms are optically pumped into the lowest magnetically
trappable state \linebreak $|\,F=1,\;m_F=-1\rangle$ and
subsequently loaded into a magnetic quadrupole trap. The trap is
generated by running 80$\,$A through the MOT coils leading to a
magnetic field gradient of $B_z' = 130\,$G/cm, which is sufficient
to hold the atoms against gravity. The number of the atoms now
amounts to nearly 1$\times$10$^9$.

\subsection{Magnetic transport and QUIC trap}

\begin{figure}
\includegraphics[width=9cm]{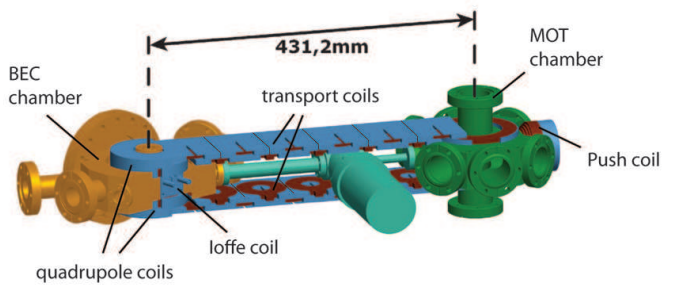}
\caption{Magnetic transport line: The neutral atom cloud is transported over a distance of 431.2$\,$mm from the MOT chamber (green) to the BEC chamber (golden). Together with their respective aluminum housings (blue) the required magnetic field coils (brown) are mounted to the stainless steel chambers.}
\label{magnetictransportline}
\end{figure}

Following the concept described in Ref.~\onlinecite{Gre2001} (see also \onlinecite{The2004}), the cold atom cloud is
transported magnetically from the MOT chamber to the BEC chamber.
For the transport we employ 13 pairs of anti-Helmholtz coils and
an additional ``push coil". By properly ramping the currents
through the coils, it is possible to smoothly shift the position
of the atoms, which are trapped in the magnetic field minimum. We
are able to move the cloud over a distance of 431$\,$mm within
1.5$\,$s. The final particle number after transport is typically
5$\times$10$^8$, corresponding to an overall transport efficiency
of about 50$\%$. The temperature of the atom cloud increases from
initially 150$\,$\textmu K to about 230$\,$\textmu K.

At the end of the transport the atoms are loaded into the QUIC
trap, which consists of the so-called
quadrupole coils together with a small diameter end coil (Ioffe
coil) (see Fig.$\,$\ref{becchamber} and \ref{opttrans}). In a first
step, the current through the quadrupole coils is ramped from
16$\,$A (used for the transport) to 36$\,$A and thus the magnetic
field gradient is increased from $B_z' = 130\,$G/cm to $B_z' =
320\,$G/cm. Subsequently, we ramp the current through the Ioffe
coil from 0 to 36$\,$A. At the end of the ramp, a single power
supply is used to drive the quadrupole coils and the Ioffe coil,
which are connected in series. Having all the QUIC coils wired up
in the same circuit minimizes heating of the atom sample and leads
to a 1/e lifetime of thermal atom clouds of about 2$\,$min. The
coil system generates an offset magnetic field of about 2$\,$G and
a nearly harmonic potential with trapping frequencies of
($\omega_x$, $\omega_y$, $\omega_z$) = 2$\pi\times$(105, 105,
20)$\,$Hz, where the z-direction is along the Ioffe axis.

We perform rf-induced forced evaporative cooling to reduce the
temperature of the atom cloud by more than two orders of
magnitude. To selectively remove hot atoms, a small coil with 3
turns and a diameter of about 20$\,$mm is placed inside the vacuum
at the bottom of the BEC chamber at a distance of 13$\,$mm from
the atoms. The coil is driven with 30$\,$dBm of rf power and the
frequency is ramped from initially 60$\,$MHz down to about
3$\,$MHz within 20$\,$s. With this procedure we are in principle
able to produce Bose-Einstein condensates of up to 3$\times 10^5$
atoms. However, for our experiments we stop the evaporation before
we reach BEC, resulting in a thermal atom cloud with about
1$\,$\textmu K temperature and an atom number of about
2$\times$10$^6$. As compared to the BEC the thermal cloud
experiences much smaller losses due to three-body collisions
during the subsequent optical transport into the science chamber.

\subsection{Optical transport of ultracold atoms}

\begin{figure}
\includegraphics[width=6.0cm]{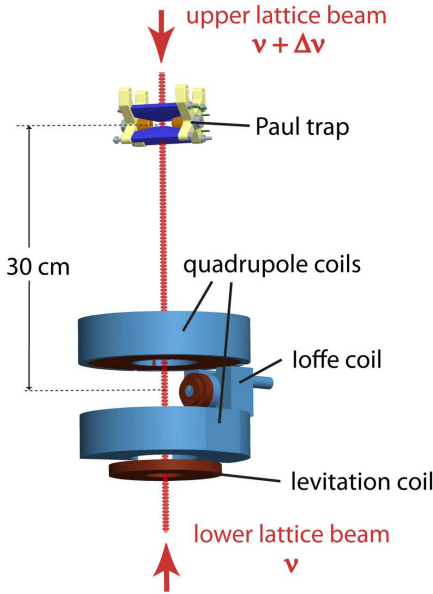}
\caption{A moving optical standing wave is used to
transport the ultracold atoms vertically from the QUIC trap in the
BEC chamber into ion trap in the science chamber. The distance between QUIC trap and Paul trap is not to scale.}
\label{opttrans}
\end{figure}

One of the key features of the experimental setup  is the vertical
long-distance optical transport of the ultracold atoms from the
BEC chamber into the science chamber. An illustration is given in
Fig.$\,$\ref{opttrans}. We follow a scheme similar to that
described in Ref.~\onlinecite{Sch2006} (where, however, the
optical transport was in horizontal direction). The ultracold
atoms are first adiabatically loaded from the QUIC trap into a
vertical far red-detuned 1-d optical lattice within 300$\,$ms. As
the lattice is set into motion it drags along the atoms, like an
elevator. After a transport distance of about 30$\,$cm,
corresponding to the distance between BEC chamber and science
chamber, the lattice is brought to a halt and the atoms are
transferred into a crossed dipole trap.

In order to load the atomic cloud into the 1-d optical lattice, it
first has to be shifted from its location close to the Ioffe coil
(where the evaporation takes place) back to the center of the
quadrupole coils which is about 5$\,$mm away. This shift is
controlled via magnetic fields from various magnetic coils. Motion
along the Ioffe axis (see Fig.$\,$\ref{becchamber}) can be induced
by changing the current through the quadrupole coils while keeping
the Ioffe current constant. For position changes along the imaging
direction (which is orthogonal to the Ioffe axis), we operate the
last pair of the magnetic transport coils. The exact position
along the (third) vertical axis is irrelevant, since the atoms can
be loaded into any antinode of the optical lattice. Nevertheless,
we added a levitation coil (see Fig.$\,$\ref{opttrans}) to the
system, which can be used to control the vertical position of the
atoms. This way we can prevent the atoms from leaving the region
to which we have good optical access. In order to optimize the
overlap between the magnetic trap and the optical standing wave,
we perform Bragg diffraction of the magnetically confined atom
cloud using the 1-d optical lattice. We adjust the position of the
QUIC trap such that the Bragg diffraction and thus the overlap is
maximal. We can determine the lattice depth experienced by the
atoms by measuring how the diffraction pattern changes as we vary
the length of the Bragg diffraction pulses. For the experimental
parameters used in our setup about 100 lattice cells are occupied
by the atoms.

\begin{figure}
\includegraphics[width=8.5cm]{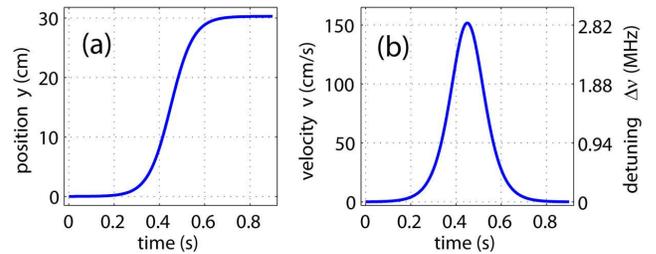}
\caption{Optical transport of ultracold atoms.  (a) A ramp of the
form $y(t)=D[\tanh(n(2t-T)/T)+\tanh(n)]/2\tanh(n)$ is chosen for
the vertical position of the atoms versus time, where $D=304\,$mm
is the transport distance, $T=0.9\,$s the transport time, and
$n=4.5$ the form parameter. (b) From the ramp $y(t)$ we can derive
the velocity $v(t)$ of the atoms and the corresponding relative
detuning (frequency shift) $\Delta\nu(t)$ between the two lattice
beams.}
\label{opttrans_xv}
\end{figure}

The optical lattice is formed by two counterpropagating collimated
Gaussian laser beams at $\lambda = 1064\,$nm with a power of
1.25$\,$W and 0.5$\,$W, respectively. For both lattice beams, the
diameter at the waist was chosen to be about 500$\,$\textmu m, so
that the divergence of the beams can be neglected. The light is
derived from a home-made fiber amplifier which is seeded by a
diode-pumped solid-state laser. \cite{Mephisto} Due to its very
low spectral linewidth ($\approx$1$\,$kHz), the laser is well
suited for the generation of an optical lattice. Both beams are
sent through acousto-optic modulators (AOM) in order to control
their frequency as well as their intensity. At the beginning of
the transport, both AOMs are driven with a rf of 80$\,$MHz. For
the transport scheme to work, it is essential that both rfs are
 phase locked to each other throughout the entire transport
sequence because sudden phase jumps would in general lead to
atomic loss. Therefore, the rf signals are generated using digital
synthesizers (AD9854) which can be locked to the same external
reference oscillator. In order to make the standing wave pattern
move with a velocity $v = \Delta\nu\,\lambda/2$, we detune the
frequency of the upper lattice beam by $\Delta\nu$. When the drive
frequency of an AOM is modified, the diffraction angle and the
beam path of the laser beam changes. To preserve the alignment of
the lattice throughout the transport, we couple the upper lattice
beam through an optical fiber before sending it to the experiment.
The fiber coupling limits the power of the upper lattice beam to
about 0.5\,W. The lattice beams enter and exit the vacuum system
through AR-coated viewports, which are attached to the chamber at
an angle of about 4$^\circ$ with respect to the (vertical)
transport axis. By this means we ensure that the reflections off
the viewports do not interfere with the standing wave.

Due to the large waist of the laser beams, the confinement of the
atoms in the optical lattice sites is more than two orders of
magnitude stronger in the axial (transport) direction than in the
radial direction. The strong axial confinement prevents gravity
from pulling the atoms out of the lattice potential even for
moderate laser intensities \cite{footnote1}. Even in the presence
of weak heating during transport, the temperature of the atom
cloud stays below 1$\,$\textmu K. This is due to evaporative
cooling from the lattice potentials which are only several
$k_\textrm{B} \times$ \textmu K  deep .

The ramp for the relative frequency shift $\Delta\nu(t)$ between
the two lattice beams is derived from the ramp for the vertical
atom position $y(t)$. Both quantities as well as the velocity
$v(t)$ are plotted in Fig.$\,$\ref{opttrans_xv} as a function of
time. During transport, the AOM frequencies and thus also the
value for the detuning $\Delta\nu(t)$ are updated every
40$\,$\textmu s. For a given transport distance $D$, the transport
time $T$ and the form parameter $n$ (see
Fig.$\,$\ref{opttrans_xv}) are optimized for maximum transport
efficiency. In our case we have $D=304\,$mm and find a maximum
efficiency of 60$\%$ for $T=0.9\,$s and $n=4.5$, ending up with
1.5$\times$10$^6$ Rb atoms in the science chamber.

The optical transport may be extended to even larger distances
than described above. As a proof of principle, we have transported
the atom cloud from the BEC chamber over 45$\,$cm to the very top
of the science chamber and then back into the BEC chamber again.
In this experiment, the total roundtrip distance of 90$\,$cm was
only limited by the length of our vacuum apparatus.

\subsection{Adjusting the lattice transport distance} 
\label{adjustD}

\begin{figure}
\includegraphics[width=8.5cm]{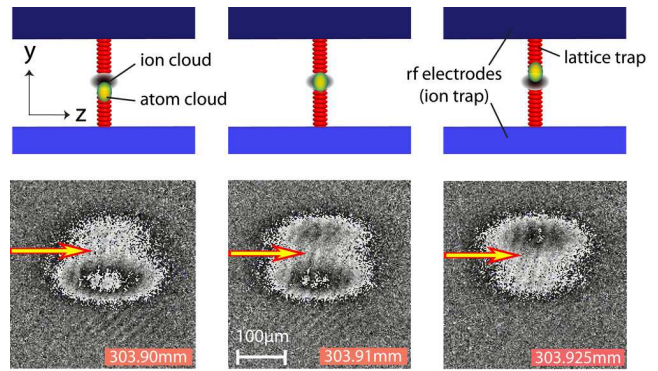}
\caption{Absorption images of the atom cloud after the interaction
with a localized ion cloud. The atoms are transported to the ion trap and brought into contact with the ion cloud for 1$\,$s. The Paul trap is loaded with a cloud of hundreds of Rb$^+$ ions, which are
responsible for the localized loss of atoms around the ion trap
center. Outside of the ion trap center the atom loss is very
small, as the atoms, being trapped in lattice sites, cannot get
into contact with the ion cloud. The transport distance is varied
between 303.90$\,$mm to 303.925$\,$mm. Arrows indicate where the
ion cloud has depleted the neutral atomic ensemble. For a
transport distance of 303.91$\,$mm the depletion region is located
in the center of the atomic cloud, indicating a good vertical
alignment. The pictures are taken after a time-of-flight of
12$\,$ms.} \label{roughalign}
\end{figure}

The atom transport has to be adjusted such that it stops at
the exact location of the trapped ion. A simple method for this would be
to use the same camera to image the locations of the ion and the
lattice held atomic cloud and to adjust the transport such that
they coincide. However, in our setup this is not possible since we
image ions and atoms with two different cameras. We thus use a
different approach which proceeds in two steps.
 First, an approximate value for the transport distance $D$ is
found by determining the position of the atom cloud relative to
the rf electrodes via standard absorption imaging. We choose $D$
such that the atoms end up roughly in the center of the ion trap
midway in between the lower and the upper electrodes.

Second, for a more precise adjustment, we load a cloud of
hundreds of ions into the Paul trap. Afterwards we transport a
freshly prepared atom cloud to the ion cloud such that the two
species get into physical contact. The ion cloud locally depletes
the atom cloud which is measured by taking an absorption
image of the atom cloud after 1$\,$s of interaction time (see Fig.$\,$\ref{roughalign}).
The depletion is due to elastic collisions
between atoms and ions in which atoms simply get kicked out
of the shallow lattice potential and are lost.\cite{Sch2010a}  The loss is local due to the
strong confinement in the lattice sites which prevents atoms outside the ion cloud from
reaching the ions. The atom cloud typically has an extension on the order
of 100$\,$\textmu m along the direction of the transport while the ion cloud is well localized within a few tens of \textmu m. In order to center the atom cloud on the ion cloud, we adjust the transport distance $D$ such that the location of loss (see arrow in Fig.$\,$\ref{roughalign}) is centered on the atomic cloud. All three images shown in Fig.$\,$\ref{roughalign} have been taken with the same ion cloud which can be repeatedly used for many depletion experiments.

\subsection{Loading of the crossed dipole trap and evaporative cooling}

\begin{figure}
\includegraphics[width=8cm]{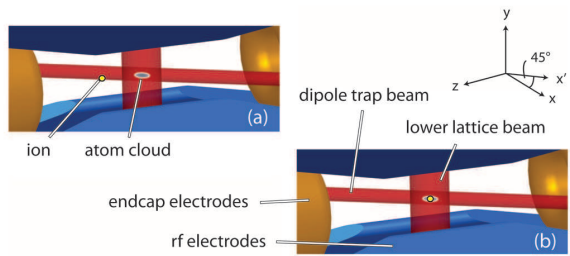}
\caption{Probing the position of the optical dipole trap with a
single ion. (a) First the ion is moved away 300$\,$\textmu
m from its normal position to prevent any collisions with the
atoms during the final evaporation stage. This is done by changing one of the endcap voltages of the
ion trap which moves the ion along the ion-trap axis (z-axis). The
position of the optical dipole trap is controlled by moving the
laser beams with the help of AOMs.
 (b) By switching back to the original endcap voltages, the ion is moved back within several ms
 to its original position where it can now probe the local density of the
 atomic cloud.} \label{overlapping}
\end{figure}

After transport, the atoms are loaded into a crossed optical dipole
trap, formed by the lower lattice beam and an additional
horizontal dipole-trap beam. This additional trapping beam is
derived from the same laser as the lattice beams and propagates
horizontally along the x'-axis (see Fig.$\,$\ref{overlapping}), which
is at an angle of 45$^\circ$ with respect to the x-axis.  It has a waist
of 50$\,$\textmu m and is ramped up to a power of 1$\,$W within
1$\,$s. Subsequently, the power of the upper lattice beam is
ramped down to zero within 1$\,$s and a crossed optical dipole
trap is formed. We are able to load about 50$\%$ of the atoms from
the 1-d optical lattice into this crossed dipole trap while
keeping the temperature of the sample below 1$\,$\textmu K.

In the next step, the depth of the dipole trap is lowered  within
4$\,$s to evaporatively cool the atoms and to end up with a BEC of
typically 10$^5$ atoms. If we prefer to work with a thermal atom
cloud ($T=100\,$nK), the evaporation is stopped immediately before
the onset of Bose-Einstein condensation. The lifetime of a thermal
atom cloud is typically on the order of 10$^2\,$s.

\subsection{Fine alignment of the crossed dipole trap}

We control the position of the crossed dipole trap dynamically
with the help of AOMs (see Fig.$\,$\ref{overlapping}). For a typical AOM with a
center frequency of 80$\,$MHz the corresponding Bragg angle is
about 10$\,$mrad. Since the frequency bandwidth is on the order of 10$\%$, the diffraction angle can be varied by about
1$\,$mrad. For distances between the AOMs and the science chamber on the order of 1$\,$m,
this results in a shift of the ion trap position by up to 1$\,$mm.

\begin{figure}
\includegraphics[width=8cm]{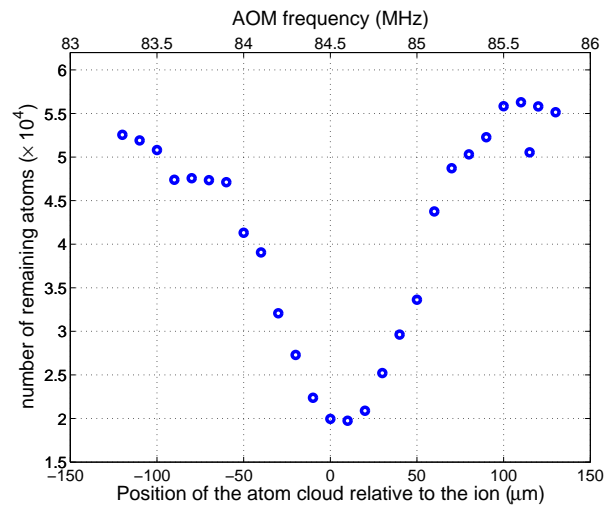}
\caption{The number of remaining atoms after a given atom-ion interaction time of $t=1\,$s. When the center of the atom cloud (i.e. the
dipole trap) coincides with the position of the ion the number of remaining atoms is minimal (see also
Ref.~\onlinecite{Sch2010a}. The position of the dipole trap is
controlled by the drive frequency of AOMs. (Here the vertical beam
of the crossed dipole trap is moved along the x'-direction.)}
\label{alignmentdata}
\end{figure}

The relative position of the dipole trap relative to the ion trap
can be accurately measured by again looking at atomic collisional
losses as previously described in section \ref{adjustD}. However,
this time we use a single ion (see Fig.$\,$\ref{overlapping}). In
order to precisely control the interaction time of the ion with
the atomic cloud, the ion is at first positioned about
300$\,$\textmu m away from its proper location so that it cannot
interact with the atoms. This is done by changing one of the
endcap voltages of the linear ion trap. This moves the ion along
the axis of the ion trap (z-axis). Afterwards, the atom cloud is
transported in from the BEC chamber and positioned by smoothly
ramping the AOM frequencies. By quickly switching back the endcap
voltages to their proper values, the ion moves back to its
original position within 2$\,$ms and starts colliding with the
atoms, kicking them out of the trap.
 This procedure is illustrated in Fig.$\,$\ref{overlapping}.
We then measure the atom loss as a function of the dipole trap
position (i.e. the AOM frequencies). A typical curve is shown in
Fig.$\,$\ref{alignmentdata} for an interaction time of about
$1\,$s, working with a thermal cloud of initially 5.5$\times10^4$
atoms. Here, the same ion is used for all data points. (The
micromotion of the ion was compensated to about 10$\,$mK for these
measurements.) The atom number reaches a minimum when the dipole
trap is centered onto the ion. We can use this measurement to
accurately overlap atom and ion trap along all three directions in
space.

\section{Summary}

In conclusion, we have designed a hybrid vacuum apparatus which
allows for cold collision experiments between ultracold neutral
atoms and trapped, cold, single ions. A main feature of this
apparatus is the spatial separation of production of the ultracold
atomic cloud and the location where experiments take place, i.e.
the linear ion trap.
In order to transport the atoms to the ion trap, we use a novel efficient optical elevator
consisting of a moving optical lattice which bridges a vertical
gap of 30$\,$cm within 1.5$\,$s. The use of a vertical transport also
leads to a non-standard design of the BEC apparatus which
houses a QUIC trap within a steel chamber. The overall layout of
the apparatus is quite modular which has advantages in terms of
stability and versatility, such that extensions can be
added to the apparatus easily in the future. Also it helped to optimize
the optical access to the region where ion-atom collision
experiments take place. In addition to discussing the overall design we also
describe various alignment and optimization procedures, e.g.
accurate positioning of the atomic cloud onto the ion trap. First
experiments with the apparatus\cite{Sch2010a} show promise for
exciting research prospects in the future.

We would like to thank Rudi Grimm for generous support during the built-up phase.  We are
grateful to Michael Drewsen and the group of Rainer Blatt for
advice on the design of the ion trap. We thank Wolfgang Limmer for
thorough proofreading of the manuscript and help in editing. We
thank Dennis Huss, Artjom Kr\"ukow, Andreas Brunner and Wolfgang Schnitzler for several technical
improvements of the set-up. This work was supported by the Austrian Science
Foundation (FWF) and the DFG within the SFB/ TR21. S.S.
acknowledges support from the Austrian Academy of Sciences within
the DOC doctorial research fellowship program.


\begin{thebibliography}{39}

\expandafter\ifx\csname natexlab\endcsname\relax\def\natexlab#1{#1}\fi
\expandafter\ifx\csname bibnamefont\endcsname\relax
 \def\bibnamefont#1{#1}\fi
\expandafter\ifx\csname bibfnamefont\endcsname\relax
  \def\bibfnamefont#1{#1}\fi
\expandafter\ifx\csname citenamefont\endcsname\relax
  \def\citenamefont#1{#1}\fi
\expandafter\ifx\csname url\endcsname\relax
  \def\url#1{\texttt{#1}}\fi
\expandafter\ifx\csname urlprefix\endcsname\relax\def\urlprefix{URL }\fi
\providecommand{\bibinfo}[2]{#2}
\providecommand{\eprint}[2][]{\url{#2}}



\bibitem[{\citenamefont{Leibfried et~al.}(2003)\citenamefont{Leibfried, Blatt, Monroe, Wineland}}]{Lei2003}
  \bibinfo{author}{\bibfnamefont{D.}~\bibnamefont{Leibfried}},
  \bibinfo{author}{\bibfnamefont{R.}~\bibnamefont{Blatt}},
  \bibinfo{author}{\bibfnamefont{C.}~\bibnamefont{Monroe}}, \bibnamefont{and}
  \bibinfo{author}{\bibfnamefont{D.}~\bibnamefont{Wineland}}, \bibinfo{journal}{Rev. Mod. Phys.}
  \textbf{\bibinfo{volume}{75}}, \bibinfo{pages}{281} (\bibinfo{year}{2003}).

\bibitem[{\citenamefont{BECoverview}(2010)\citenamefont{BECoverview}}]{BEC2002}
  \bibinfo{author}{\bibfnamefont{}~\bibnamefont{Nature Insight on Ultracold Matter}}, \bibinfo{journal}{Nature}
  \textbf{\bibinfo{volume}{416}}, \bibinfo{pages}{205-246} (\bibinfo{year}{2002}).

  \bibitem[{\citenamefont{Hawley et~al.}(1991)\citenamefont{Hawley, Smith}}]{Haw1991}
  \bibinfo{author}{\bibfnamefont{M.}~\bibnamefont{Hawley}},  \bibnamefont{and}
  \bibinfo{author}{\bibfnamefont{M.~A.}~\bibnamefont{Smith}}, \bibinfo{journal}{J. Chem. Phys.}
  \textbf{\bibinfo{volume}{95}}, \bibinfo{pages}{8662} (\bibinfo{year}{1991}).

\bibitem[{\citenamefont{Gerlich et~al.}(2002)\citenamefont{Gerlich, Herbst, Roue}}]{Ger2002}
  \bibinfo{author}{\bibfnamefont{D.}~\bibnamefont{Gerlich}},
  \bibinfo{author}{\bibfnamefont{E.}~\bibnamefont{Herbst}}, \bibnamefont{and}
  \bibinfo{author}{\bibfnamefont{E.}~\bibnamefont{Roue}}, \bibinfo{journal}{Planetary and Space Science}
  \textbf{\bibinfo{volume}{50}}, \bibinfo{pages}{1275} (\bibinfo{year}{2002}).

  \bibitem[{\citenamefont{Gerlich}(2005)\citenamefont{Gerlich}}]{Ger2005}
  \bibinfo{author}{\bibfnamefont{D.}~\bibnamefont{Gerlich}}, \bibinfo{journal}{Phys. Chem. Chem. Phys.}
  \textbf{\bibinfo{volume}{7}}, \bibinfo{pages}{1583} (\bibinfo{year}{2005}).

\bibitem{Ott08}
    R. Otto, J. Mikosch, S. Trippel, M. Weidemüller, and R. Wester
    Phys. Rev. Lett. {\bf 101}, 063201 (2008).

\bibitem[{\citenamefont{Willitsch et~al.}(2008)\citenamefont{Willitsch, Bell, Gingell, Procter, Softley}}]{Wil2008}
  \bibinfo{author}{\bibfnamefont{S.}~\bibnamefont{Willitsch}},
  \bibinfo{author}{\bibfnamefont{M.~T.}~\bibnamefont{Bell}},
  \bibinfo{author}{\bibfnamefont{A.~D.}~\bibnamefont{Gingell}},
  \bibinfo{author}{\bibfnamefont{S.~R.}~\bibnamefont{Procter}}, \bibnamefont{and}
  \bibinfo{author}{\bibfnamefont{T.~P.}~\bibnamefont{Softley}}, \bibinfo{journal}{Phys. Rev. Lett.}
  \textbf{\bibinfo{volume}{100}}, \bibinfo{pages}{043203} (\bibinfo{year}{2008}).

\bibitem[{\citenamefont{Cetina et~al.}(2007)\citenamefont{Cetina, Grier, Campbell, Chuang, Vuletic}}]{Cet2007}
  \bibinfo{author}{\bibfnamefont{M.}~\bibnamefont{Cetina}},
  \bibinfo{author}{\bibfnamefont{A.}~\bibnamefont{Grier}},
  \bibinfo{author}{\bibfnamefont{J.}~\bibnamefont{Campbell}},
  \bibinfo{author}{\bibfnamefont{I.}~\bibnamefont{Chuang}},  \bibnamefont{and}
  \bibinfo{author}{\bibfnamefont{V.}~\bibnamefont{Vuletic}}, \bibinfo{journal}{Phys. Rev. A}
  \textbf{\bibinfo{volume}{76}}, \bibinfo{pages}{041401} (\bibinfo{year}{2007}).

\bibitem[{\citenamefont{Grier et~al.}(2009)\citenamefont{Grier, Cetina, Orucevic, Vuletic}}]{Gri2009}
  \bibinfo{author}{\bibfnamefont{A.~T.}~\bibnamefont{Grier}},
  \bibinfo{author}{\bibfnamefont{M.}~\bibnamefont{Cetina}},
  \bibinfo{author}{\bibfnamefont{F.}~\bibnamefont{Orucevic}}, \bibnamefont{and}
  \bibinfo{author}{\bibfnamefont{V.}~\bibnamefont{Vuletic}}, \bibinfo{journal}{Phys. Rev. Lett.}
  \textbf{\bibinfo{volume}{102}}, \bibinfo{pages}{223201} (\bibinfo{year}{2009}).

\bibitem{Hal11} F.~H.~J. Hall, M. Aymar, N. Bouloufa-Maafa, O.
    Dulieu, and S. Willitsch, Phys. Rev. Lett. {\bf 107}, 243202
    (2011).

\bibitem{Rel11} W.~G. Rellergert, S.~T. Sullivan, S. Kotochigova, A. Petrov, K. Chen, S.~J. Schowalter, and E.~R. Hudson, Phys. Rev. Lett. {\bf 107}, 243201 (2011).

\bibitem{Rav11}
    K. Ravi, S. Lee, A. Sharma, G. Werth, S.A. Rangwala,
    Appl. Phys. B: Lasers and Optics, DOI 10.1007/s00340-011-4726-6
    (2011).

\bibitem{WeidWest} M. Weidem\"uller and Roland Wester, private communication.

\bibitem[{\citenamefont{Zipkes et~al.}(2010)\citenamefont{Zipkes, Palzer, Sias, K\"{o}hl}}]{Zip2010}
  \bibinfo{author}{\bibfnamefont{C.}~\bibnamefont{Zipkes}},
  \bibinfo{author}{\bibfnamefont{S.}~\bibnamefont{Palzer}},
  \bibinfo{author}{\bibfnamefont{C.}~\bibnamefont{Sias}}, \bibnamefont{and}
  \bibinfo{author}{\bibfnamefont{M.}~\bibnamefont{K\"{o}hl}},
  \bibinfo{journal}{Nature} \textbf{\bibinfo{volume}{464}}, \bibinfo{pages}{388-391}
  (\bibinfo{year}{2010}).

\bibitem[{\citenamefont{Zipkes et~al.}(2010)\citenamefont{Zipkes, Ratschbacher, Palzer, Sias, K\"{o}hl}}]{Zip2010a}
  \bibinfo{author}{\bibfnamefont{C.}~\bibnamefont{Zipkes}},
  \bibinfo{author}{\bibfnamefont{S.}~\bibnamefont{Palzer}},
  \bibinfo{author}{\bibfnamefont{L.}~\bibnamefont{Ratschbacher}},
  \bibinfo{author}{\bibfnamefont{C.}~\bibnamefont{Sias}}, \bibnamefont{and}
  \bibinfo{author}{\bibfnamefont{M.}~\bibnamefont{K\"{o}hl}}, \bibinfo{journal}{Phys. Rev. Lett.}
  \textbf{\bibinfo{volume}{105}}, \bibinfo{pages}{133201} (\bibinfo{year}{2010}).

\bibitem[{\citenamefont{Zipkes et~al.}(2011)\citenamefont{Zipkes, Ratschbacher, Sias, K\"{o}hl}}]{Zip2011}
  \bibinfo{author}{\bibfnamefont{C.}~\bibnamefont{Zipkes}},
  \bibinfo{author}{\bibfnamefont{L.}~\bibnamefont{Ratschbacher}},
  \bibinfo{author}{\bibfnamefont{S.}~\bibnamefont{Palzer}},
  \bibinfo{author}{\bibfnamefont{C.}~\bibnamefont{Sias}}, \bibnamefont{and}
  \bibinfo{author}{\bibfnamefont{M.}~\bibnamefont{K\"{o}hl}}, \bibinfo{journal}{J. Phys.: Conf. Ser.}
  \textbf{\bibinfo{volume}{264}}, \bibinfo{pages}{012019} (\bibinfo{year}{2011}).


\bibitem[{\citenamefont{Schmid et~al.}(2010)\citenamefont{Schmid, H\"arter, Hecker Denschlag}}]{Sch2010a}
  \bibinfo{author}{\bibfnamefont{S.}~\bibnamefont{Schmid}},
  \bibinfo{author}{\bibfnamefont{A.}~\bibnamefont{H\"arter}}, \bibnamefont{and}
  \bibinfo{author}{\bibfnamefont{J.}~\bibnamefont{Hecker Denschlag}}, \bibinfo{journal}{Phys. Rev. Lett.}
  \textbf{\bibinfo{volume}{105}}, \bibinfo{pages}{133202} (\bibinfo{year}{2010}).

\bibitem[{\citenamefont{C\^{o}t\'{e} et~al.}(2000)\citenamefont{C\^{o}t\'{e}, Dalgarno}}]{Cot2000b}
  \bibinfo{author}{\bibfnamefont{R.}~\bibnamefont{C\^{o}t\'{e}}}, \bibnamefont{and}
  \bibinfo{author}{\bibfnamefont{A}~\bibnamefont{Dalgarno}}, \bibinfo{journal}{Phys. Rev. A}
  \textbf{\bibinfo{volume}{62}}, \bibinfo{pages}{012709} (\bibinfo{year}{2000}).

\bibitem[{\citenamefont{Idziaszek et~al.}(2009)\citenamefont{Idziaszek, Calarco, Julienne, Simoni}}]{Idz2009}
  \bibinfo{author}{\bibfnamefont{Z.}~\bibnamefont{Idziaszek}},
  \bibinfo{author}{\bibfnamefont{T.}~\bibnamefont{Calarco}},
  \bibinfo{author}{\bibfnamefont{P.~S.}~\bibnamefont{Julienne}}, \bibnamefont{and}
  \bibinfo{author}{\bibfnamefont{A.}~\bibnamefont{Simoni}},  \bibinfo{journal}{Phys. Rev. A}
  \textbf{\bibinfo{volume}{79}}, \bibinfo{pages}{010702} (\bibinfo{year}{2009}).

  \bibitem[{\citenamefont{Staanum et~al.}(2010)\citenamefont{Staanum, Hojbjerre, Skyt, Hansen, Drewsen}}]{Sta2010}
  \bibinfo{author}{\bibfnamefont{P.~F.}~\bibnamefont{Staanum}},
  \bibinfo{author}{\bibfnamefont{K.}~\bibnamefont{Hojbjerre}},
  \bibinfo{author}{\bibfnamefont{P.}~\bibnamefont{Skyt}},
  \bibinfo{author}{\bibfnamefont{A.}~\bibnamefont{Hansen}}, \bibnamefont{and}
  \bibinfo{author}{\bibfnamefont{M.}~\bibnamefont{Drewsen}}, \bibinfo{journal}{Nat. Phys.}
  \textbf{\bibinfo{volume}{6}}, \bibinfo{pages}{271} (\bibinfo{year}{2010}).

\bibitem[{\citenamefont{Schneider et~al.}(2010)\citenamefont{Schneider, Roth, Duncker, Ernsting, Schiller}}]{Sch2010b}
  \bibinfo{author}{\bibfnamefont{T.}~\bibnamefont{Schneider}},
  \bibinfo{author}{\bibfnamefont{B.}~\bibnamefont{Roth}},
  \bibinfo{author}{\bibfnamefont{H.}~\bibnamefont{Duncker}},
  \bibinfo{author}{\bibfnamefont{I.}~\bibnamefont{Ernsting}}, \bibnamefont{and}
  \bibinfo{author}{\bibfnamefont{S.}~\bibnamefont{Schiller}}, \bibinfo{journal}{Nat. Phys.}
  \textbf{\bibinfo{volume}{6}}, \bibinfo{pages}{275} (\bibinfo{year}{2010}).

\bibitem[{\citenamefont{Massignan et~al.}(2005)\citenamefont{Massignan, Pethick, Smith}}]{Mas2005}
  \bibinfo{author}{\bibfnamefont{P.}~\bibnamefont{Massignan}},
  \bibinfo{author}{\bibfnamefont{C.~J.}~\bibnamefont{Pethick}}, \bibnamefont{and}
  \bibinfo{author}{\bibfnamefont{H.}~\bibnamefont{Smith}},  \bibinfo{journal}{Phys. Rev. A}
  \textbf{\bibinfo{volume}{71}}, \bibinfo{pages}{023606} (\bibinfo{year}{2005}).

\bibitem[{\citenamefont{Kalas et~al.}(2006)\citenamefont{Kalas, Blume}}]{Kal2006}
  \bibinfo{author}{\bibfnamefont{R.~M.}~\bibnamefont{Kalas}}, \bibnamefont{and}
  \bibinfo{author}{\bibfnamefont{D.}~\bibnamefont{Blume}},  \bibinfo{journal}{Phys. Rev. A}
  \textbf{\bibinfo{volume}{73}}, \bibinfo{pages}{043608} (\bibinfo{year}{2006}).

\bibitem[{\citenamefont{Cucchietti et~al.}(2006)\citenamefont{Cucchietti, Timmermans}}]{Cuc2006}
  \bibinfo{author}{\bibfnamefont{F.~M.}~\bibnamefont{Cucchietti}}, \bibnamefont{and}
  \bibinfo{author}{\bibfnamefont{E.}~\bibnamefont{Timmermans}},  \bibinfo{journal}{Phys. Rev. Lett.}
  \textbf{\bibinfo{volume}{96}}, \bibinfo{pages}{210401} (\bibinfo{year}{2006}).

\bibitem[{\citenamefont{C\^{o}t\'{e}}(2000)\citenamefont{C\^{o}t\'{e}}}]{Cot2000a}
  \bibinfo{author}{\bibfnamefont{R.}~\bibnamefont{C\^{o}t\'{e}}}, \bibinfo{journal}{Phys. Rev. Lett.}
  \textbf{\bibinfo{volume}{85}}, \bibinfo{pages}{5316} (\bibinfo{year}{2000}).

\bibitem[{\citenamefont{Cote et~al.}(2002)\citenamefont{C\^{o}t\'{e}, Kharchenko, Lukin}}]{Cot2002}
  \bibinfo{author}{\bibfnamefont{R.}~\bibnamefont{C\^{o}t\'{e}}},
  \bibinfo{author}{\bibfnamefont{V.}~\bibnamefont{Kharchenko}}, \bibnamefont{and}
  \bibinfo{author}{\bibfnamefont{M.~D.}~\bibnamefont{Lukin}}, \bibinfo{journal}{Phys. Rev. Lett.}
  \textbf{\bibinfo{volume}{89}}, \bibinfo{pages}{093001} (\bibinfo{year}{2002}).

\bibitem[{\citenamefont{igp}()}]{igp}
  \bibinfo{author}{\bibnamefont{Varian StarCell 75$\,$l/s}}.

\bibitem[{\citenamefont{gauge}()}]{gauge}
  \bibinfo{author}{\bibnamefont{Varian UHV-24p}}.

\bibitem[{\citenamefont{Rb}()}]{Rb}
  \bibinfo{author}{\bibnamefont{Sigma-Aldrich Part-No 276332-1G}}.

\bibitem[{\citenamefont{TiSub}()}]{TiSub}
  \bibinfo{author}{\bibnamefont{Titanium Sublimation pump TSP (filament type) from Varian}}.

\bibitem[{\citenamefont{Esslinger et~al.}(1998)\citenamefont, Bloch, H\"ansch}]{Ess1998}
  \bibinfo{author}{\bibfnamefont{T.}~\bibnamefont{Esslinger}},
  \bibinfo{author}{\bibfnamefont{I.}~\bibnamefont{Bloch}}, \bibnamefont{and}
  \bibinfo{author}{\bibfnamefont{T.~W.}~\bibnamefont{H\"ansch}}, \bibinfo{journal}{Phys. Rev. A}
  \textbf{\bibinfo{volume}{58}}, \bibinfo{pages}{R2664-R2667} (\bibinfo{year}{1998}).

\bibitem[{\citenamefont{channel}()}]{channel}
  \bibinfo{author}{\bibnamefont{CEM-4823G from Burle}}.

\bibitem[{\citenamefont{Ravi et~al.}(1992)\citenamefont{Raizen}}]{Rav2011}
  \bibinfo{author}{\bibfnamefont{K.}~\bibnamefont{Ravi}},
  \bibinfo{author}{\bibfnamefont{S.}~\bibnamefont{Lee}},
  \bibinfo{author}{\bibfnamefont{A.}~\bibnamefont{Sharma}},
  \bibinfo{author}{\bibfnamefont{G.}~\bibnamefont{Werth}},   \bibnamefont{and}
  \bibinfo{author}{\bibfnamefont{S.~A.}~\bibnamefont{Rangwala}}, \bibinfo{journal}{arXiv:1009.4515v3}
  (\bibinfo{year}{2011}).

\bibitem[{\citenamefont{Raizen et~al.}(1992)\citenamefont{Raizen}}]{Rai1992}
  \bibinfo{author}{\bibfnamefont{M.}~\bibnamefont{Raizen}},
  \bibinfo{author}{\bibfnamefont{J.~M.}~\bibnamefont{Gilligan}},
  \bibinfo{author}{\bibfnamefont{J.~C.}~\bibnamefont{Bergquist}},
  \bibinfo{author}{\bibfnamefont{W.~M.}~\bibnamefont{Itano}},   \bibnamefont{and}
  \bibinfo{author}{\bibfnamefont{D.~J.}~\bibnamefont{Wineland}}, \bibinfo{journal}{Phys. Rev. A}
  \textbf{\bibinfo{volume}{45}}, \bibinfo{pages}{6493} (\bibinfo{year}{1992}).

\bibitem[{\citenamefont{Agilent}()}]{Agilent}
  \bibinfo{author}{\bibnamefont{Agilent 33220A}}.

\bibitem[{\citenamefont{Minicircuits5W}()}]{Minicircuits5W}
  \bibinfo{author}{\bibnamefont{Minicircuits ZHL-5W-1}}.

\bibitem[{\citenamefont{Berkeland et~al.}(1992)\citenamefont{Berkeland}}]{Ber1998}
  \bibinfo{author}{\bibfnamefont{D.~J.}~\bibnamefont{Berkeland}},
  \bibinfo{author}{\bibfnamefont{J.~D.}~\bibnamefont{Miller}},
  \bibinfo{author}{\bibfnamefont{J.~C.}~\bibnamefont{Bergquist}},
  \bibinfo{author}{\bibfnamefont{W.~M.}~\bibnamefont{Itano}},   \bibnamefont{and}
  \bibinfo{author}{\bibfnamefont{D.~J.}~\bibnamefont{Wineland}}, \bibinfo{journal}{J. App. Phys.}
  \textbf{\bibinfo{volume}{83}}, \bibinfo{pages}{10} (\bibinfo{year}{1998}).

\bibitem[{\citenamefont{Baoven}()}]{Baoven}
  \bibinfo{author}{\bibnamefont{Alvasource from the company Alvatec}}.

\bibitem[{\citenamefont{413er}()}]{413er}
  \bibinfo{author}{\bibnamefont{DL-100 from Toptica}}.

\bibitem{Dre1983} R.~W.~P. Drever, J.~L. Hall, F.~V. Kowalski, J. Hough, G.~M. Ford, A.~J. Munley und H. Ward, Appl. Phys. B {\bf 31}, 97-105
    (1983).

\bibitem{Bla2001} E.~D. Black, Am. J. Phys. {\bf 69}, 79
    (2001).

\bibitem{Ber1998} D.~J. Berkeland, J.~D. Miller, J.~C. Bergquist, W.~M. Itano, D.~J. Wineland, J. Appl. Phys. {\bf 83}, 5025
    (1998).

\bibitem[{\citenamefont{McCarron et~al.}(2008)\citenamefont{McCarron, King, Cornish}}]{McC2008}
  \bibinfo{author}{\bibfnamefont{D.~J.}~\bibnamefont{McCarron}},
  \bibinfo{author}{\bibfnamefont{S.~A.}~\bibnamefont{King}}, \bibnamefont{and}
  \bibinfo{author}{\bibfnamefont{S.~L.}~\bibnamefont{Cornish}}, \bibinfo{journal}{Meas. Sci. Technol.}
  \textbf{\bibinfo{volume}{19}}, \bibinfo{pages}{105601} (\bibinfo{year}{2008}).

\bibitem[{\citenamefont{BoosTA}()}]{BoosTA}
  \bibinfo{author}{\bibnamefont{BoosTA from Toptica}}.

\bibitem[{\citenamefont{Bjorklund et~al.}(1983)\citenamefont{Bjorklund, Levenson, Lenth, Ortiz}}]{Bjo1983}
  \bibinfo{author}{\bibfnamefont{G.~C.}~\bibnamefont{Bjorklund}},
  \bibinfo{author}{\bibfnamefont{M.~D.}~\bibnamefont{Levenson}},
  \bibinfo{author}{\bibfnamefont{W.}~\bibnamefont{Lenth}}, \bibnamefont{and}
  \bibinfo{author}{\bibfnamefont{C.}~\bibnamefont{Ortiz}},  \bibinfo{journal}{Appl. Phys. B}
  \textbf{\bibinfo{volume}{32}}, \bibinfo{pages}{145-152} (\bibinfo{year}{1983}).

\bibitem[{\citenamefont{Greiner et~al.}(2001)\citenamefont{Greiner, Bloch, H\"ansch, Esslinger}}]{Gre2001}
  \bibinfo{author}{\bibfnamefont{M.}~\bibnamefont{Greiner}},
  \bibinfo{author}{\bibfnamefont{I.}~\bibnamefont{Bloch}},
  \bibinfo{author}{\bibfnamefont{T.~W.}~\bibnamefont{H\"ansch}}, \bibnamefont{and}
  \bibinfo{author}{\bibfnamefont{T.}~\bibnamefont{Esslinger}}, \bibinfo{journal}{Phys. Rev. A}
  \textbf{\bibinfo{volume}{63}}, \bibinfo{pages}{031401} (\bibinfo{year}{2001}).

  \bibitem{The2004} M. Theis, G. Thalhammer, K. Winkler, M. Hellwig, G. Ruff, R. Grimm, and J. Hecker Denschlag {\bf 93}, 123001
    (2004).

\bibitem[{\citenamefont{Schmid et~al.}(2006)\citenamefont{Schmid, Thalhammer, Winkler, Lang, Hecker Denschlag}}]{Sch2006}
  \bibinfo{author}{\bibfnamefont{S.}~\bibnamefont{Schmid}},
  \bibinfo{author}{\bibfnamefont{G.}~\bibnamefont{Thalhammer}},
  \bibinfo{author}{\bibfnamefont{K.}~\bibnamefont{Winkler}},
  \bibinfo{author}{\bibfnamefont{F.}~\bibnamefont{Lang}},  \bibnamefont{and}
  \bibinfo{author}{\bibfnamefont{J.}~\bibnamefont{Hecker Denschlag}},
  \bibinfo{journal}{New. J. Phys.} \textbf{\bibinfo{volume}{8}}, \bibinfo{pages}{159}
  (\bibinfo{year}{2006}).


\bibitem[{\citenamefont{Mephisto}()}]{Mephisto}
  \bibinfo{author}{\bibnamefont{Mephisto from the company Innolight}}.

\bibitem{footnote1} This is in fact the main reason for the vacuum chamber layout where the
       position of the science chamber is located above the BEC chamber. A horizontal transfer into the science chamber would have been plagued with compensating the gravitational force to prevent atomic losses.


\end{thebibliography}
\end{document}